\documentclass[final,3p,times]{elsarticle}

\usepackage{graphics,graphicx,dcolumn,bm,fleqn,epic,eepic,float,ulem}
\usepackage{amssymb,amsmath}
\usepackage{multirow,rotate}
\usepackage{color}
\usepackage[english]{babel}

\definecolor{red}{rgb}{1,0,0}
\definecolor{green}{rgb}{0,1,0}
\definecolor{blue}{rgb}{0,0,1}

\newcommand{\f}{\mathbf}

\newcommand{\no}{NO$_2$}

\journal{}
\begin{document}
%%%%%%%%%%%%%%%%%%%%%%%%%%%%%%%%%%%%%%%%%%%%%%%%%%%%%%%%%%%%%%%%%%%%%%%%%

\begin{frontmatter}

\title{Air quality prediction using optimal neural networks with stochastic variables}

\author{Ana Russo$^{a}$}
\author{Frank Raischel$^{b}$}
\author{Pedro G.~Lind$^{b,c,d}$}

\address{$^a$Center for Geophysics , IDL, 
         University of Lisbon
         1749-016 Lisboa, Portugal}
\address{$^b$Center for Theoretical and Computational Physics, 
         University of Lisbon, Av.~Prof.~Gama Pinto 2, 
         1649-003 Lisbon, Portugal}
\address{$^c$TWIST - Turbulence, Wind energy and Stochastics,
         Institute of Physics, Carl-von-Ossietzky University of 
         Oldenburg, DE-26111 Oldenburg, Germany}
\address{$^d$ForWind - Center for Wind Energy Research, Institute of Physics,
         Carl-von-Ossietzky University of Oldenburg, DE-26111 Oldenburg, 
         Germany}

%\begin{linenumbers}
\begin{abstract} 
We apply recent methods in stochastic data analysis for discovering a set 
of few stochastic variables that represent the relevant information on a 
multivariate stochastic system, used as input for artificial neural 
networks models for  air quality forecast.
We show that using these derived variables as input variables for 
training the neural networks 
it is possible to significantly 
reduce the amount of input variables necessary for the neural network 
model, without considerably changing the predictive power of the model.
The reduced set of variables including these derived variables is therefore 
proposed as optimal variable set for training neural networks models in 
forecasting geophysical and weather properties. 
Finally, we briefly discuss other possible applications of such optimized 
neural network models.
\end{abstract}

%%%%PACS e Keywords
%\end{linenumbers}
\end{frontmatter}

%%%%%%%%%%%
%\begin{linenumbers}
\section{Introduction}
\label{sec:intro}

Urban air pollution is a complex mixture of toxic components
with considerable impact on the inhabitants of urban regions, particularly those
belonging to sensitive groups, such as children and people with 
previous heart and respiratory insufficiency\cite{kolehmainen2001}.
Therefore, forecasting the temporal evolution of air pollution concentrations in specific 
urban locations emerges as a priority for guaranteeing life quality in 
urban and metropolitan centers. 
With this aim, in order to identify and predict in advance episodes 
of low air quality at regional and local scales, air quality forecasting 
models have been developed, considering the characteristics of atmospheric 
pollution and its consequent impact on people’s health and life quality.
%%%%%%%%%%%%%%%%%%%%%%%%%%%%%%%%%%%%%%%%%%%%%%%%%%%%%%%%%%%%%%%%%%%%
\begin{figure}[t]
  \centering
  \includegraphics[width=0.5\textwidth]{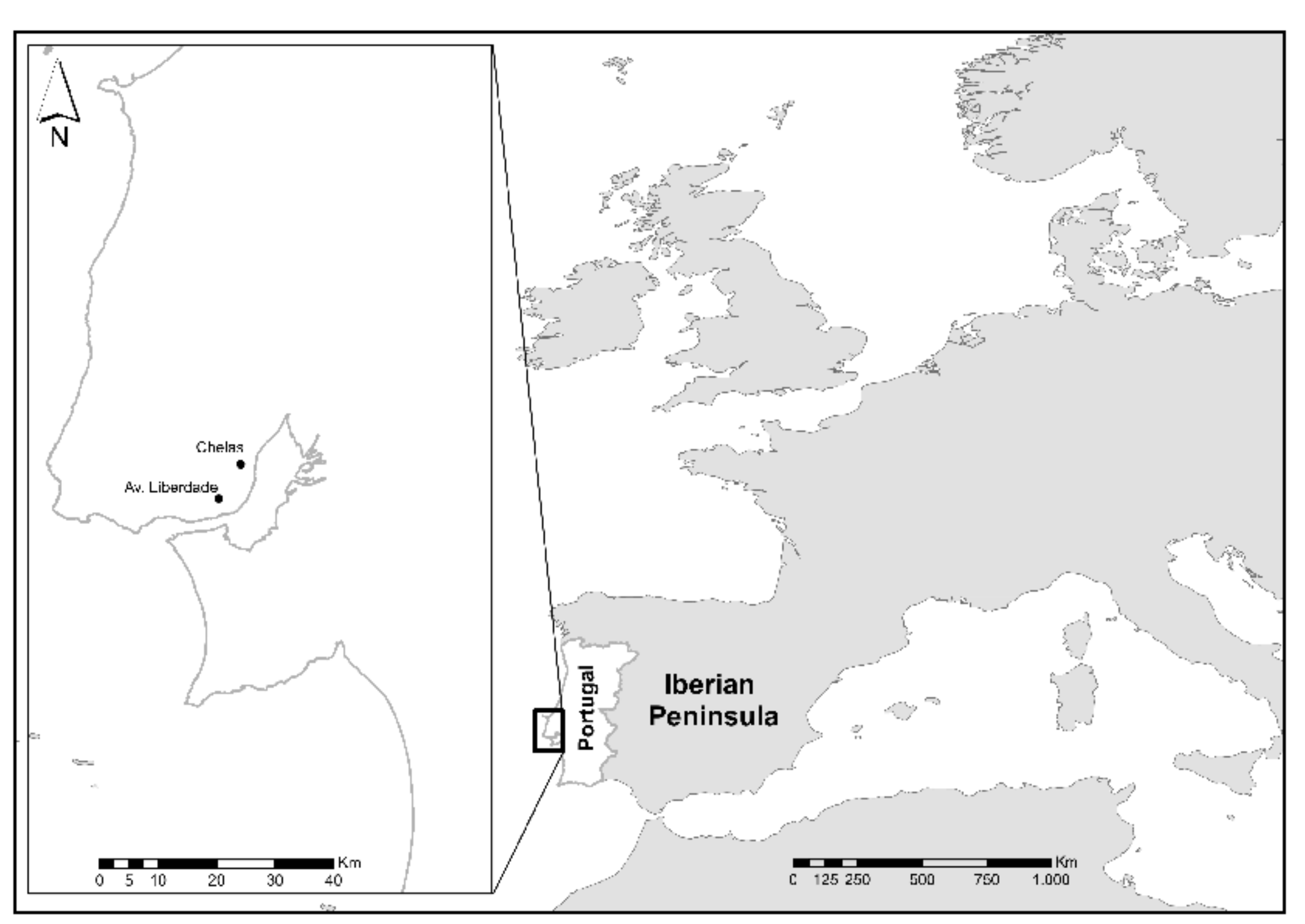}
  \caption{\protect 
           \no\ measurement stations in the region of Lisbon (Portugal)
           at the Southwestern coast of Europe.
           In this paper we focus on the set of measurements taken at the
           stations of Chelas and Avenida da Liberdade, 
           approximately $4$ km distant apart, with
           approximately $3x10^4$ data points. Each data set is extracted 
           within the period between 2002 and 2006, with a frequency
           of $1$ hour$^{-1}$.} 
\label{fig1}
\end{figure}
%%%%%%%%%%%%%%%%%%%%%%%%%%%%%%%%%%%%%%%%%%%%%%%%%%%%%%%%%%%%%%%%%%%%

Straightforward approaches such as Box models\cite{middleton1997}, Gaussian 
plume models\cite{reich1999}, persistence and regression models\cite{shi1999} 
are commonly applied to characterize and forecast air pollutants’ dispersion. 
These models are easy to implement and allow for the rapid calculation of forecasts. However, 
they  include significant simplifications\cite{luecken2006} and usually do not 
describe the processes and interactions that control the transport and 
chemical behavior of pollutants in the atmosphere\cite{luecken2006}, 
important for instance for secondary pollutants\cite{sokhi2006}. 
Improvements have been made with deterministic dispersion models 
and statistical-based approaches, which however, being 
highly non–linear\cite{binbhu2012}, require a large amount of 
accurate input data and are considerably expensive from the computational 
point of view\cite{dutot}.

A promising alternative to all these models are artificial neural networks 
(ANN)\cite{binbhu2012,nejadkoorki2012,gardner1997}. 
Several ANN models have already been used for air quality forecast, in particular
for forecasting hourly averages\cite{kolehmainen2001,perez2000,kukkonen2003} 
and daily maxima\cite{perez2002}. 
Further, several authors compared already the potential of different 
approaches when applied to different pollutants and prediction time 
lags\cite{kukkonen2003,YiPrybutok1996,GardnerDorling2000,Hooyberghs2005}. 
Still, though successful in many situations and having considerably less 
restrictions on the input data,  large training data sets are usually 
required to improve accuracy and minimize 
uncertainty in the output data, which up to now  has been a significant disadvantage of these models.

Recently, we applied methods from stochastic data analysis and statistical 
physics for deriving variables with reduced stochastic 
fluctuations\cite{vitor} to empirical data in sets of \no\ concentration 
measurements\cite{no2}. Such methods were introduced in the late 
nineties\cite{friedrich97,physrepreview} for analyzing measurements on 
complex stochastic processes, aiming for a quantitative estimation of drift and diffusion 
functions from sets of measurements that fully define the evolution 
equation of the underlying stochastic variables.
The framework has already been  applied successfully,  for instance to describe 
turbulent flows\cite{friedrich97} and the evolution of climate 
indices\cite{lind05,lind07}, performance curves of wind turbines\cite{wind},
stock market indices\cite{friedrich00}, 
and oil prices\cite{ghasemi07}. 
At the same time, the basic method has been refined in particular for
data with low sampling frequency\cite{kleinhans05,lade09} 
and subjected to strong measurement noise\cite{boettcher06,lind10,carvalho2010}.

In this paper we present an important application of such variables:
using them as input for training ANN 
enables one to reduce considerably the amount of 
input data needed for achieving a given accuracy.
We argue that this reduction in the number of input variables is possible 
because the derived variables incorporate temporal correlations between 
independent and spatially separated monitoring stations.
Moreover, as we quantitatively show below, 
when using this reduced amount of information that includes the derived
variables, the predictive power of the ANN is not significantly
changed, which is a major advantage when working with observational data which might include missing values.
Combining a faster ANN training with the same predictive 
power may improve the ability and capability of alert system for air quality
in large urban centers.
We start in Sec.~\ref{sec:methods} by briefly describing ANN models
as well as the main points of the stochastic data analysis procedure
used. In Sec.~\ref{sec:data} the empirical data is described, comprising
two different data sets of \no\ concentration measures in the city
of Lisbon, Portugal (see Fig.~\ref{fig1}). 
In Sec.~\ref{sec:results} the results are
discussed in the light of predictive power measures,
and Sec.~\ref{sec:conclusions} concludes the paper.

%%%%%%%%%%%%%%%%%%%%%%%%%%%
\section{Methods}
\label{sec:methods}

\subsection{The Neural Network framework}

\begin{figure}[t]
  \centering
  \includegraphics[width=0.5\textwidth]{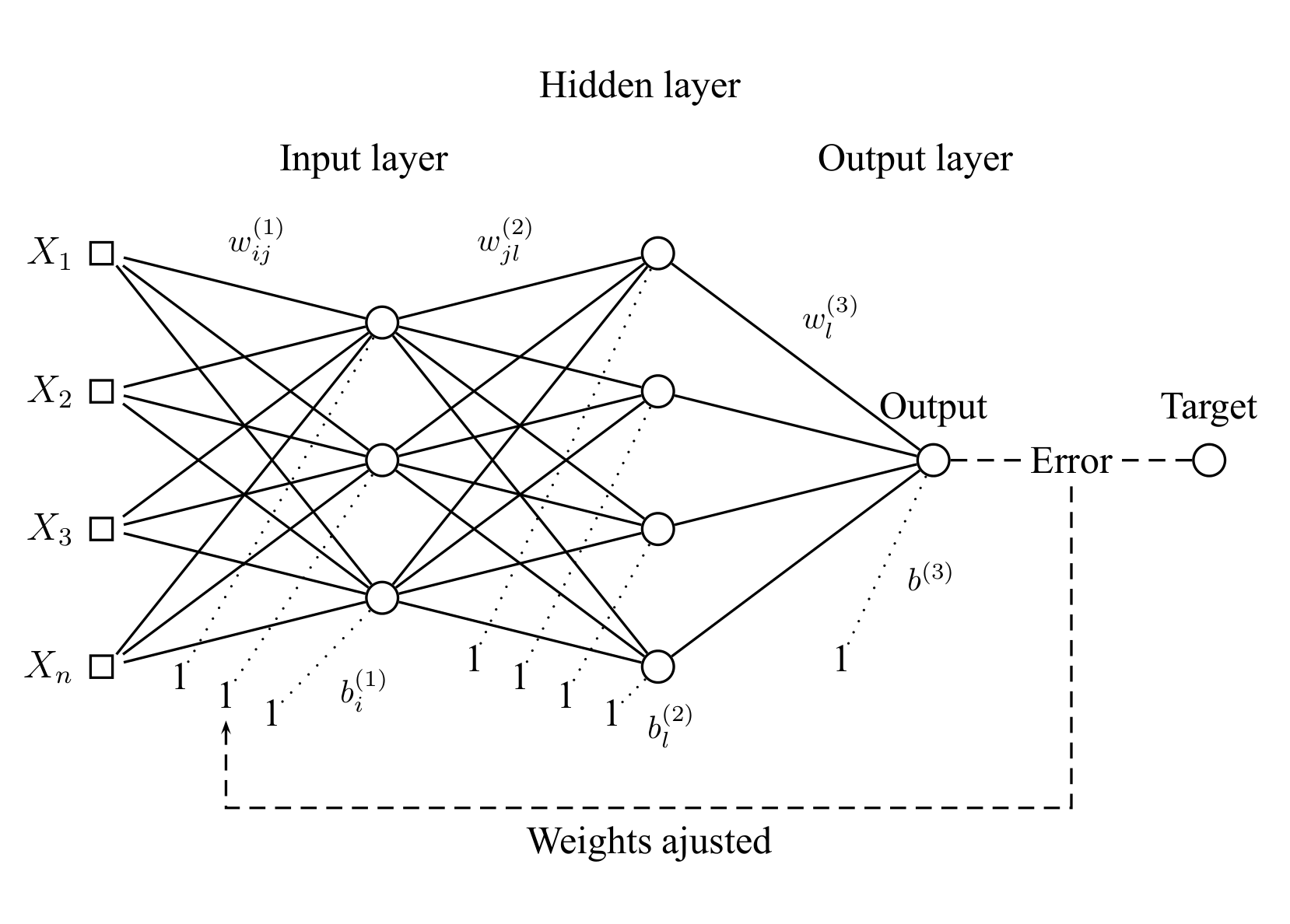}
\caption{\protect Example of the structure for a feed-forward artificial 
           neural network model with three layers. Variables $X_{i=1,...,n}$ represent 
					 the input variables (neurons), $\omega$ the weights associated to each neuron 
					 and $b$ the bias vectors which combined will produce an output within certain error limits.}
\label{fig2}
\end{figure}

Artificial neural network models are mathematical models 
inspired by the functioning of  nervous 
systems\cite{gardner1997,Cobourn2000,Agirre-Basurko2006}, 
which are composed by a number of interconnected entities, the 
artificial neurons (see Fig.~\ref{fig2}).

These neurons may be associated in many different 
ways\cite{Agirre-Basurko2006,Haykin1999}, depending on the characteristics 
of the proposed problem. 
To construct an ANN model the air pollution system is considered as a system 
that receives information from $n$ distinct sets of inputs $X_i$ 
($i=1,\dots,n$), namely weather parameters and air pollution properties, and 
produces a specific output, in our case the concentration of the \no\ pollutant\cite{gardner1997}. 
No prior knowledge about the relationship between input and output 
variables is assumed.
The input variables should be independent from each other and each one
is represented by its own input neuron $i=1,\dots,n$.
Each neuron computes a linear combination of the weighted inputs $\omega_{ij}$,
including a bias term  $b_{i}$, from the links feeding into it and the corresponding
summed value $C_j=\sum_i \omega_{ij}X_i + b_{i}$ is transformed using a  
function $f$, either linear or non-linear such as log-sigmoid or
hyperbolic tangent. The bias term is included in order to allow the activation functions to be offset from zero and it can be set randomly or set to a desired value (e.g. dummy input with a magnitude equal to 1).  
The output obtained is then passed as a new input 
$\tilde{X}_j = f(C_j)$ to other nodes in the following layer, 
usually named hidden layer. %, as shown in Fig.~\ref{fig2}. 
Though one is allowed to use several neurons in this hidden layer,
it is generally advantageous to somehow minimize the number of hidden 
neurons, in order to improve the generalization capabilities of the model 
and also to avoid over-fitting. 
In particular, a simple one-layer ANN structure with just one neuron 
employing a linear activation function reduces to the well-known
linear regression model\cite{Weisberg1985}. 
The ANN models used here are based on a feed-forward configuration of 
the multilayer perceptron that has been
used by several authors \citep{Hooyberghs2005,Papanastasiou}. 
A large number of architectures has been tested. 
The use of two layers was verified to be 
sufficient. The use of more layers was concluded to be redundant, and 
therefore we only use two layers, one input layer and one hidden layer.
%%%%%%%%%%%%%%%%%%%%%%%%%%%%%%%%%%%%%%%%%%%%%%%%%%%%%%%%%%%%%%%%%%%%
%%%%%%%%%%%%%%%%%%%%%%%%%%%%%%%%%%%%%%%%%%%%%%%%%%%%%%%%%%%%%%%%%%%%

Having such a framework of input variables  and sets of functions, the ANN has to be 
trained in order to obtain the best estimate for each weight $\omega$. 
The weight values are determined by an optimization procedure, the so-called
learning algorithm\cite{Haykin1999}, which in our case uses a cross-validation 
procedure\cite{wilks2006} to ensure stability of the model. 
The cross-validation was applied dividing the available period into four sets and completing the calibration-validation procedure four times independently, i.e., from the 4 years of data available, data from 3 years were used to build the model and data from 1 year for validation. The validation year was then cycled through the 4 year period. For each validation year, a set of performance measures were computed between the observed real values and the ANN forecasts, namely, the Pearson correlation coeﬃcient (PC); the root mean square error (RMSE); and the skill against persistence (Sp). The outcomes of each resulting performance measure were analyzed for each year and then averaged for the complete period, resulting on an average value for each monitoring station.

There are several learning algorithms, depending on whether the ANN
model is linear or non-linear.
For linear ANN models, the learning algorithm is typically based on the 
Widrow-Hoff learning-rule, also known as the least mean square rule, 
which produces a unique solution corresponding to the absolute minimum 
value of the error surface\cite{TrigoPalutikof1999}.
For non-linear models, the back-propagation (BP) is one of the most popular and common training procedures used, which is described in depth in the literature \cite{Haykin1999,TrigoPalutikof1999}. It has been shown in literature that BP training algorithms have two caveats: convergence may be slow and the final weights may be trapped in local minima over the highly complex error surface \cite{TrigoPalutikof1999}. As an alternative, the 
Levenberg-Marquardt method\cite{numrecip} minimizes an error 
function in ``damped'' procedures, i.e.~selecting steps proportional 
to the gradient of the error function.
The Levenberg-Marquardt method requires more 
memory \cite{Haykin1999,TrigoPalutikof1999} than
the Widrow-Hoff rule, but has the advantage of converging faster and 
with a higher effective robustness than most BP 
schemes\cite{TrigoPalutikof1999} because it avoids having to compute 
second-order derivatives. The Levenberg-Marquardt method was appliedfor this study.
%The non-linear ANN models' performance should be compared to other models' performances. 
Together with the persistence model, which is the simplest way of producing a forecast and assumes that the conditions at the time of the forecast will not change, i.e., the forecast for each time step simply corresponds to the value of the previous time step, the linear regression model constitutes the baseline against which the performance of non-linear 
ANN models are usually compared. Due to a certain level of memory that characterizes air pollutants, persistence corresponds to a benchmark model considerably more difficult to beat than climatology \cite{demuzere}. 

%%%%%%%%%%%%%%%%%%%%%%%%%%%%%%%%%%%%%%%%%%%%%%%%%%%%%%%%%%%%%%%%%%%%%%%%%%%%%
\subsection{Deriving optimal stochastic variables}
\label{subsec:optimal}

In this section we briefly describe how from a number $K$ of sets of 
measurements, one is able to derive a set of few  stochastic variables containing 
information from all of them.

This procedure assumes that corresponding to each  measurement variable 
there is a property that evolves according to some stochastic equation.
More precisely, the $K$-dimensional state vector  
${\bf X}=(x_1,...,x_K)$ characterizes the set of $K$ measurements at each
time-step and evolves according to the It\^o-Langevin 
equations\cite{fpeq,gard}:
\begin{equation}
\frac{d \mathbf{X}}{d t}= 
                     \mathbf{h}(\mathbf{X})
                     + \mathbf{g}(\mathbf{X}) 
                     \mathbf{\Gamma}(t) , 
\label{Lang2DVect}
\end{equation}
where $\mathbf{\Gamma}=(\Gamma_1,\dots,\Gamma_K)$ is a set of $K$ independent
stochastic forces with Gaussian distribution fulfilling 
$\langle \Gamma_i(t)\rangle = 0$ and 
$\langle \Gamma_i(t)\Gamma_j(t')\rangle = 2\delta_{ij}\delta(t-t')$.
On the right hand side of Eq.~(\ref{Lang2DVect})
the term with function $\mathbf{h}=\{ h_i\}$
describes the determinist part, which drifts the system,
while the term with $\mathbf{g}=\{ g_{ij} \}$
account for the amplitude of the stochastic contributions characterized
through the properties of 
$\mathbf{\Gamma}$\cite{physrepreview}.

The heart of the method lies in the fact that the coefficients 
$\mathbf{h}$ and  $\mathbf{g}$ are closely related to the drift vectors 
and diffusion matrices describing the evolution of the joint probability 
density function of the vector state $\mathbf{X}$ by means of the corresponding 
Fokker-Planck equation\cite{fpeq,gard}.
As has been shown previously in other 
contexts\cite{physrepreview,lind05,friedrich00,ghasemi07,kleinhans05,%
boettcher06,lind10}, the drift vector and the diffusion matrix can 
therefore be extracted directly from the data set $\mathbf{Y}$\cite{no2} 
\begin{subequations}
\begin{eqnarray}
D_i^{(1)}(\mathbf{X}) &=& h_i (\mathbf{X}) \cr
                     &\approx&
           \lim_{\tau \rightarrow0}\frac{1}{\tau}
           \left\langle Y_i(t+\tau)-Y_i(t) | 
           \mathbf{Y}(t)=\mathbf{X} \right \rangle \label{M1}\\
D^{(2)}_{ij}(\mathbf{X}) &=& \sum^K_{k=1}g_{ik}(\mathbf{X})g_{jk}
                           (\mathbf{X}) \cr
           &\approx&
           \lim_{\tau \rightarrow 0}\frac{1}{2\tau}
        \left\langle (Y_i(t+\tau)-Y_i(t)) 
        \right. \nonumber \\
        && \left. 
        \cdot (Y_j(t+\tau)-Y_j(t)) | 
        \mathbf{Y}(t)=\mathbf{X}  \right\rangle \label{M2}
\end{eqnarray}
\label{DefCoefKM}
\end{subequations}
for $i,j=1,\dots,K$ and where 
$\langle \cdot| {\mathbf{Y}(t)=\mathbf{X}} \rangle$ symbolizes conditional 
averaging over all measurements $\mathbf{Y}(t)$ that fulfill the condition 
$\mathbf{Y}(t)=\mathbf{X}$. 

The last equation in both Eqs.~(\ref{M1}) and (\ref{M2}) yields
the operational definition of the first and second 
conditional moments\cite{physrepreview,lind10}, respectively. 
The limit in Eq.~(\ref{DefCoefKM}) 
is typically  approximated by the slope of a linear fit of the 
corresponding conditional moments at small $\tau$. 
When this linear fit is not possible, as it is the case for the \no\ 
measurements under consideration, an alternative estimate\cite{kleinhans05} 
is to consider the first value of $M(\tau)/\tau$ at the lowest value
of $\tau$.
Additional analysis has been  carried out\cite{no2}, namely confirming the 
Markovian properties of the data sets, which is a prerequisite for assuming 
a  Langevin process, Eq.~\eqref{Lang2DVect}. In case Markovian properties 
are not observed, it should be noted that this method may be still applied 
with alternative procedures\cite{boettcher06,lind10,carvalho2010}. 
Furthermore, in case a Fourier spectrum shows periodicities, those should 
be filtered out by a proper detrending procedure.
More details about how to apply Eqs.~(\ref{DefCoefKM}) can be found
in Ref.~\cite{no2}.

We apply this framework to the two-dimensional system of detrended  \no\ 
concentration measurements taken in two  stations in Lisbon situated in Chelas and Avenida
da Liberdade. We consider therefore a vector ${\bf X}=(x_1,x_2)$.
Both sets of measurements do exhibit Markovian properties\cite{no2},
and therefore both drift and diffusion functions are properly derived.

To arrive to the optimal variables, we next determine the eigensystem 
of the diffusion matrix and investigate its principal directions\cite{vitor,%
gradisek_eigenvectors,vanMourik_eigenvectors}. These principal directions
are computed for each mesh point previously defined in phase space\cite{no2}.
The aim is to obtain the transform of the original coordinates 
$\mathbf{X}=\{ x_i \}$ into new ones $\mathbf{\tilde{X}}=\{ \tilde{x}_i \}$, 
such that the the diffusion matrix is diagonalized. 
The transformation $\f{\tilde{X}}={\f{F}}(\f{X},t)$ is a two-times 
continuously differentiable function.
In general,
the eigenvalues of the diffusion matrix indicate the amplitude of 
the stochastic force and the corresponding eigenvector indicates 
the phase space direction towards which such force acts. 

In this way, the stochastic contribution is decoupled for each new
variable. Consequently, if the eigenvalues in the transformed 
coordinates are significantly different, we are able to restrict our 
investigation to the coordinates with lower stochastic sources, i.e.~lower
eigenvalues. 
For such eigenvalues, the vector field of their eigenvectors 
defines the path in phase space towards which the fluctuations are
minimal.
Additionally, if these $j$ eigenvalues are very small compared to all the 
others, the corresponding stochastic forces can be neglected and the system 
can be assumed to have only $K-j$ independent stochastic forces,
reducing the number of stochastic variables in the system.
For all the details  see Ref.~\cite{no2}.
%%%%%%%%%%%%%%%%%%%%%%%%%%%%%%%%%%%%%%%%%%%%%%%%%%%%%%%%%%%%%%%%%%%%
\begin{figure}[tb]
  \centering
  \includegraphics[width=0.5\textwidth]{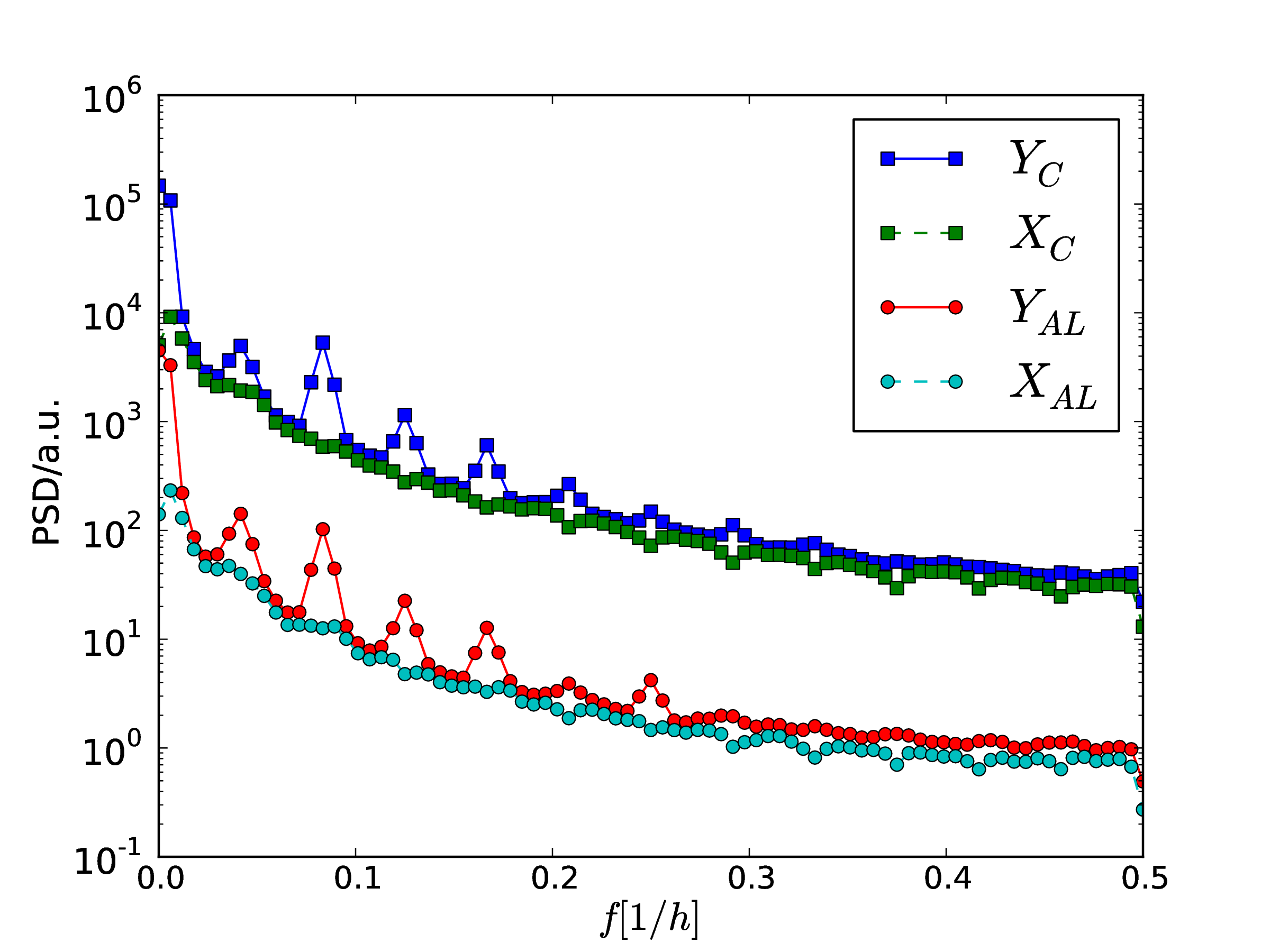}
  \caption{\protect 
           Fourier spectrum (power spectral density PSD) of the \no\ 
           measurements in Chelas and 
           Avenida da Liberdade stations. While the original 
           measurements, symbolized by $Y_C$ and $Y_{AL}$, 
           show some periodic modes resulting from the daily, three-day 
           and weekly cycles (see text), the detrended variables, 
           $X_C$ and $X_{AL}$, both in Chelas and Avenida da 
           Liberdade have these periodicities suppressed. 
           Plots are vertically offset for better visibility. }
\label{fig3}
\end{figure}
%%%%%%%%%%%%%%%%%%%%%%%%%%%%%%%%%%%%%%%%%%%%%%%%%%%%%%%%%%%%%%%%%%%%

%%%%%%%%%%%
\section{Data}
\label{sec:data}

\subsection{Target data}
\label{sec:targetdata}

We consider hourly measurements of \no\ concentrations in the metropolitan 
region of Lisbon, Portugal, namely at the monitoring stations of Chelas (C) and
Avenida da Liberdade (AL) (see Fig.~\ref{fig1}). 
The data were recorded from 
2002 to 2006, corresponding to $\sim 3x10^4$ measurement points
with roughly 1\% of discarded  values, due to incomplete or
erroneous measurements. 
The stations are located  at a distance of $ 4.8$km from each other. 
In the following, the \no\ concentrations at the stations of Chelas and 
Avenida da Liberdade will be designated as $Y_{C}(t) $ and $Y_{AL}(t)$,
respectively, omitting the temporal dependency when not necessary.
Figure \ref{fig3} shows the Fourier spectrum of both these data series. 

\subsection{Input data for ANN training}
\label{sec:inputdata}

The ANN input data sets consist of the aforementioned hourly \no\ concentration measurements, and of concentrations of two more 
pollutants, namely NO, and CO, also measured at the monitoring 
stations of Chelas and Avenida da Liberdade, from January 1st 2002 until
December 31st of 2006.

The first four years are used to construct the 
models and year 2006 is used for independent evaluation.
More specifically the prediction is done in two steps.
In the first step, we consider only the period 2002-2005. 
For this four years we 
take the first three, 2002-2004, for training the ANN and derive the 
respective parameter values of the ANN model. Using that ANN we predict 
year 2005. Then we consider 2002, 2003 and 2005 for training the ANN, and 
predict 2004. Similar procedure is done for predicting 2002 and 2003. 
In the second step, we use the parameter values obtained in the first step, namely, 
weights and biases, for predicting 2006.

Besides pollutant concentrations, we also consider daily maximum temperature, 
daily mean wind direction and speed, daily humidity, daily radiance, 
hourly mean temperature, hourly pressure and hourly relative humidity,
boundary layer height (BLH) from the European Center for Medium Weather 
Forecast (ECMWF), circulation weather type (CWT) at the regional scale 
determined for Portugal according to Trigo and DaCamara\cite{trigocamara}, 
North Atlantic Oscillation (NAO) index from NCEP-NOAA,
two weekly cycles and two yearly cycles. 
Specifically, BLH fields were retrieved from the 3 hourly ECMWF 40 years 
reanalysis (http://data-portal.ecmwf.int/data) for the 2002-2006 period. 
Afterward, we extracted the 00:00 UTC (BLH1), 03:00 UTC (BLH2), 9:00 (BLH3) 
and 21:00 UTC (BLH4) data from the retrieved BLH fields.
Together with these variables we consider the two transformed variables
derived through the procedure described in Sec.~\ref{subsec:optimal}, before 
we  suggest a method for reducing the number of input variables. In total 
there are $48$ variables that are available as input data for the ANN model 
(See Tab.~\ref{fig_tabela_var}).
%%%%%%%%%%%Tabela com as variaveis utilizadas por cada modelo%%%%%%%%%%%%%%
\begin{table}[t]
%\begin{ruledtabular}
\begin{tabular}{|c||c|c|}
\hline
\small{\bf Model} & \small{\bf Variables} & \small{\bf Time $t$ and Lag $\Delta t$}\\
\hline
\small{\textbf{$M_s$}} & \small {\no }&\small{$t-\Delta t$ to $t-24$}\\
                       & \small {NO, CO }& \small{$t-\Delta t$}\\
		       & \small {NAO index} & \small{$t$, $t-\Delta t$}\\
		       & \small {CWT} & \small{$t$, $t-1$}\\
       & \small {BLH1, BLH2, BLH3, BLH4} & \small{$t-\Delta t$}\\
       & \small {Daily maximum temperature }& \small{$t-\Delta t$}\\
       & \small {Daily mean wind direction}&\small{$t-\Delta t$}\\
       & \small {Daily mean wind speed}& \small{$t-\Delta t$}\\	
       & \small {Daily mean humidity}& \small{$t-\Delta t$}\\ 
       & \small {Daily mean radiance}& \small{$t-\Delta t$}\\ 
       & \small {Hourly mean temperature}&\small{$t-\Delta t$}\\
       & \small {Hourly mean pressure}&\small{$t-\Delta t$}\\
       & \small {Hourly mean relative humidity }&\small{$t-\Delta t$}\\
       & \small {$sin(2\pi t/365)$, $cos(2\pi t/365)$}&\small {$t$}\\
       & \small {$sin(2\pi t/7)$, $cos(2\pi t/7)$}& \small{$t$}\\
\hline
\small{\textbf{$M_{s+}$}}&\small{The same as {$M_{s}$}}&\small{(see above)}\\
                        &\small{$V_O^+$, $V_O^-$}&\small{$t-1$}\\
\hline
\end{tabular}
%\end{ruledtabular}
  \caption{\protect 
           Description of available input parameters for prediction.
           Here the time-lags are $\Delta t=1,3,6,12$ and $24$ hours.
           For model $M_s$ one considers a total maximum number of $46$ variables 
           and for model $M_{s+}$ the two transformed variables
           $V_0^+$ and $V_0^-$ are added.}
\label{fig_tabela_var}
\end{table}
%%%%%%%%%%%%%%%%%%%%%%%%%%%%%%%%%%%%%%%%%%%%%%%%%%%%%%%%%%%%%%%%%%%

The two transformed variables are obtained from the original
\no\ measurements $Y_{C}$ and $Y_{AL}$ by first detrending them.
As shown in Fig.~\ref{fig3} both sets of measurements $Y_{C}$ and
$Y_{AL}$ have periodic contributions, which must be filtered out.
These periodicities describes daily, weekly, seasonal and 
yearly variations of the concentration due to anthropogenic
routines and to periodic atmospheric processes\cite{kolehmainen2001}.
In particular, the 24 hours and one week cycles  are both traffic 
related and mirror daily and weekly cycles.
By filtering out these cycles, through a proper detrending, 
one is reduced to the stochastic contribution solely,
which can be modeled through a stochastic differential equation\cite{no2,fpeq},
having both the  deterministic forcing in the drift vector $\mathbf{h}$, and the
diffusive fluctuation, $\mathbf{g}$. 

The detrended series derived from the original measurements $Y_{C}$ and $Y_{AL}$, represented henceforth as
$X_{C}$ and $X_{AL}$, respectively, are obtained as follows.
The data is partitioned into segments of length $N$, multiple
of all relevant periodic modes. 
Our simulations have shown that two such partitions are needed
in the present case, as one partition alone is not sufficient to remove the periodicities completely.
First, averages over $N=52$ weeks are performed, and afterward a second detrending 
with $N=1$ day follows on consecutive periods of $14$ days.
Next, a mean segment is calculated by averaging measurements with the same 
relative position in the periodic segment. 
Finally, the detrended data set is obtained by subtracting the 
respective values of the mean segment from the measured data\cite{no2}.
Whereas the Markovian framework can strictly only be applied after removing all the periodicities by detrending (or a similar procedure), we have verified that the estimates of the drift and diffusion coefficients and of the mean orientation angle are not altered significantly by the second step of the detrending procedure, i.~e.~after removing the periodicities only partially, which confirms that our detrending method does not introduce artifacts. 

As indicated by Fig.~\ref{fig3}, 
the detrended data do not show patterns of periodicity, satisfying both 
Markov properties as well as the delta-correlated 
signature of the Gaussian-like noise\cite{no2}. 
One therefore may consider the series $X_C$ and $X_{AL}$ as a set described by 
two coupled Langevin Equations. 

Having obtained the detrended data, the method described in Sec.~\ref{subsec:optimal}
is then applied, yielding a two-dimensional drift vector and diffusion
matrix. The diffusion matrix is diagonalized, yielding two eigenvalues and
their normalized eigenvectors, orthogonal to each other.
Multiplying them by their corresponding eigenvalues yields two orthogonal 
vectors that define an ellipse in phase space $(X_C,X_{AL})$.
These diffusion ellipses have a specific orientation defined through an
angle whose absolute value quantifies the relative off-diagonal contribution 
that describes the coupling of the noise terms by the diffusion 
matrix\cite{no2}. 
Rotating all the ellipses by the orientation angle, aligns the largest
eigenvector along one of the coordinate axis and the smallest eigenvector
along the other one. 
At each mesh point we construct from  the numerical values of the two 
detrended time series,  and the respective functional dependency of the 
eigenvalues on the original variables, two transformed time series, 
$V_O^{+}$ and $V_0^{-}$, which
we take as additional input data for the 
ANN model. We then test the resulting improvement of the forecasts. 
The optimal variable corresponding to the largest eigenvalue is henceforth
symbolized by $V_O^+$ 
while the other will be represented by $V_O^-$.
As reported below, the variable $V_O^+$ shows a higher rank of importance
than the variable $V_O^-$ when selecting the most important variables
for ANN training.
%%%%%%%%%%%Tabela de sumario do numero de variaveis%%%%%%%%%%%%%%
\begin{table}[t]
\begin{tabular}{|c||c|c|c|c|c|c|}
\hline
 & {\bf \textbf{$\Delta t$} (h)} & {\bf {$M_s$} }& \textbf{$M_{s+}$} & \textbf{$M_{s,FSR}$} & 
\textbf{$M_{O}$} & {\bf Dev.}\\
\hline
\multirow{4}{1.0cm}{{\bf C}}   & 1  & 46 & 48 & 22 & 13 & {\bf 41\%}\\
                                 & 3  & 44 & 46 & 26 & 26 &  {\bf 0\%}\\
                                 & 6  & 41 & 43 & 31 & 15 & {\bf 55\%}\\
                                 & 12 & 36 & 38 & 28 & 7  & {\bf 75\%}\\
                                 & 24 & 27 & 29 & 19 & 2  & {\bf 89\%}\\\hline
\multirow{4}{1.0cm}{{\bf AL}}    & 1  & 46 & 48 & 29 & 13 & {\bf 52\%}\\
                                 & 3  & 44 & 46 & 25 & 18 & {\bf 28\%}\\
                                 & 6  & 41 & 43 & 28 & 13 & {\bf 54\%}\\
                                 & 12 & 36 & 38 & 27 & 10  & {\bf 63\%}\\
                                 & 24 & 27 & 29 & 20 & 2  & {\bf 90\%}\\\hline
\end{tabular}
\caption{\protect 
         Number of retained variables for each model at both $C$ and $AL$ 
         monitoring stations. Five different time lags $\Delta t$ are 
         considered. The last column indicates the relative reduction of 
         the number of input variables when substituting the standard 
         model $M_{s,FSR}$ by model $M_{O}$.}
\label{fig_tabela_y}
\end{table}
%%%%%%%%%%%%%%%%%%%%%%%%%%%%%%%%%%%%%%%%%%%%%%%%%%%%%%%%%%%%%%%%%%%

To select a reduced set of the $M_{s+}$ input variables we
conducted a forward stepwise regression (FSR)
between the meteorological and the air quality variables for each 
monitoring station independently.
The FSR allows variables to be optimized in order to predict \no\ at 
each monitoring station. This procedure starts with the variable most 
correlated with the target, and adds one new variable which, 
together with the previous one, most accurately predicts the target, 
i.e.~the variable that reduces the predicting error the most.
One adds iteratively new variables in this way, ordering the set of 
$M_{s+}$ variables by descending order of correlation with the target.
The procedure stops when any new variable does not significantly 
reduce the prediction error. The corresponding significance of error 
reduction is measured by a partial $F$-test\cite{numrecip}. 
With this approach, given an initial set of variables for one 
monitoring station, one is able to select the best subset of variables
for predicting the evolution of \no\ concentration at that station.

In this study, we consider four distinct sets of variables
for the input data, defining therefore four distinct ANN models.

The first set is the complete set of (up to, depending on the lag 
investigated) initial variables, neglecting
the transformed variables $V_O^+$ and $V_O^-$, and is used for training
the ANN, which then defines the standard ANN model, henceforth represented
by $M_s$. 
Another set 
corresponds to the first set with 
the transformed variables $V_O^+$ and $V_O^-$, yielding the model $M_{s+}$.

The third set is extracted from the $M_{s+}$ variables by FSR, 
yielding the model $M_{s,FSR}$.

The forth set is a subset of this extended set and uses the correlation 
ordering made for all the variables, where it considers only the variable $V_O^+$ and the variables
having stronger correlation with the target. This is our optimal
model $M_{O}$.

Table \ref{fig_tabela_y} shows the number of variables used for each
model and for five different time lags.
Depending on this time lag, namely $t-1$, $t-3$, $t-6$, $t-12$
 and $t-24$ hours, 
the input data includes hourly data from the previous $t=1,\dots,24$ hours, 
$t=3,\dots,24$ hours, $t=6,\dots,24$ hours, $t=12,\dots,24$ hours and $t=24$ hours, respectively. 

While model $M_{s+}$ uses always two more variables than
the standard model $M_{s}$, the number of variables for the optimal model
$M_{O}$ is typically smaller then the number of variables for $M_{s,FSR}$.

In the last column of Tab.~\ref{fig_tabela_y}, we indicate the relative 
reduction on the number of input variables when using model $M_{O}$ instead 
of model $M_{s,FSR}$.
The reduction of the number of variables used 
for training the ANN model is typically above $50\%$, with only
three exceptions, raising up to $90\%$ 
for the largest time-lag predictions.
Notice that by construction the number in column $M_{O}$ indicates the 
rank of the variable $V_O^+$ when ordering the variables according to 
their correlation with the target. 
This result also shows the strong correlation between our transformed
variable $V_O^+$ and the target.

In all cases, the variable $V_O^+$ shows a higher correlation with the
target than the variable $V_O^-$, therefore $V_O^-$ does not appear
in the subsets for $M_{O}$. The reason for this higher correlation
is associated with the strength of the corresponding fluctuations.
Since the variable $V_O^+$ is the transformed variable corresponding
to the largest eigenvalue of the diffusion matrix for the coupled
system of both \no\ concentrations, $X_C$ and $X_{AL}$, the pair of 
variables fluctuate stronger along this direction and therefore
contains a larger part of the correlations than the other transformed 
variable $V_O^-$. This situation is somehow
comparable to the assumptions of the  principal component analysis
model \cite{Pearson}.

%%%%%%%%%%%%%%%%%%%%%%%%%%%%%%%%%%%%%%%%%%%%%%%%%%%%%%%%%%%%%%%%%%%%
\begin{figure}[h!]
 \centering
 \includegraphics[trim=50 0 50 10,width=.5\textwidth]{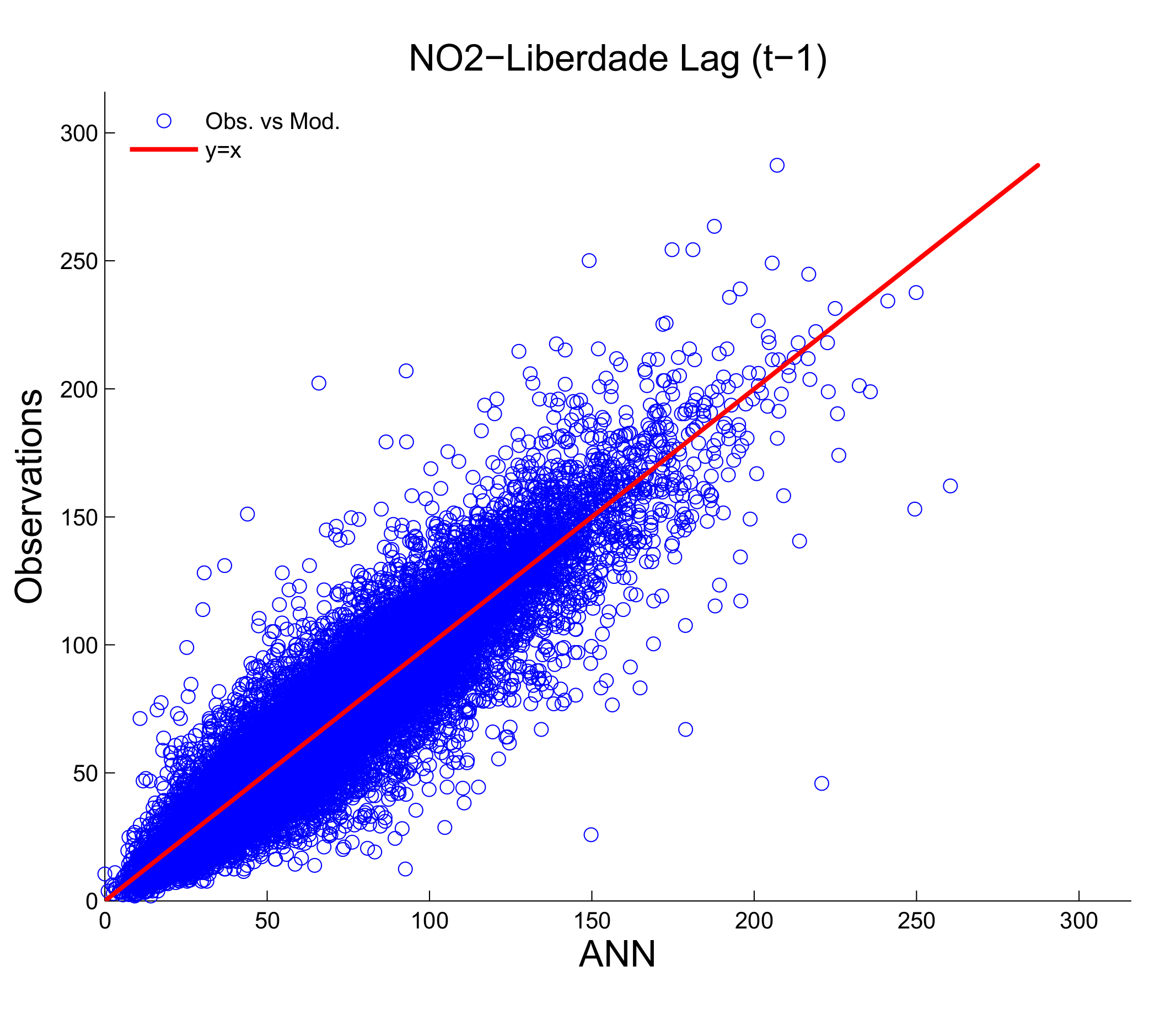}
\includegraphics[trim=50 0 50 10,width=.5\textwidth]{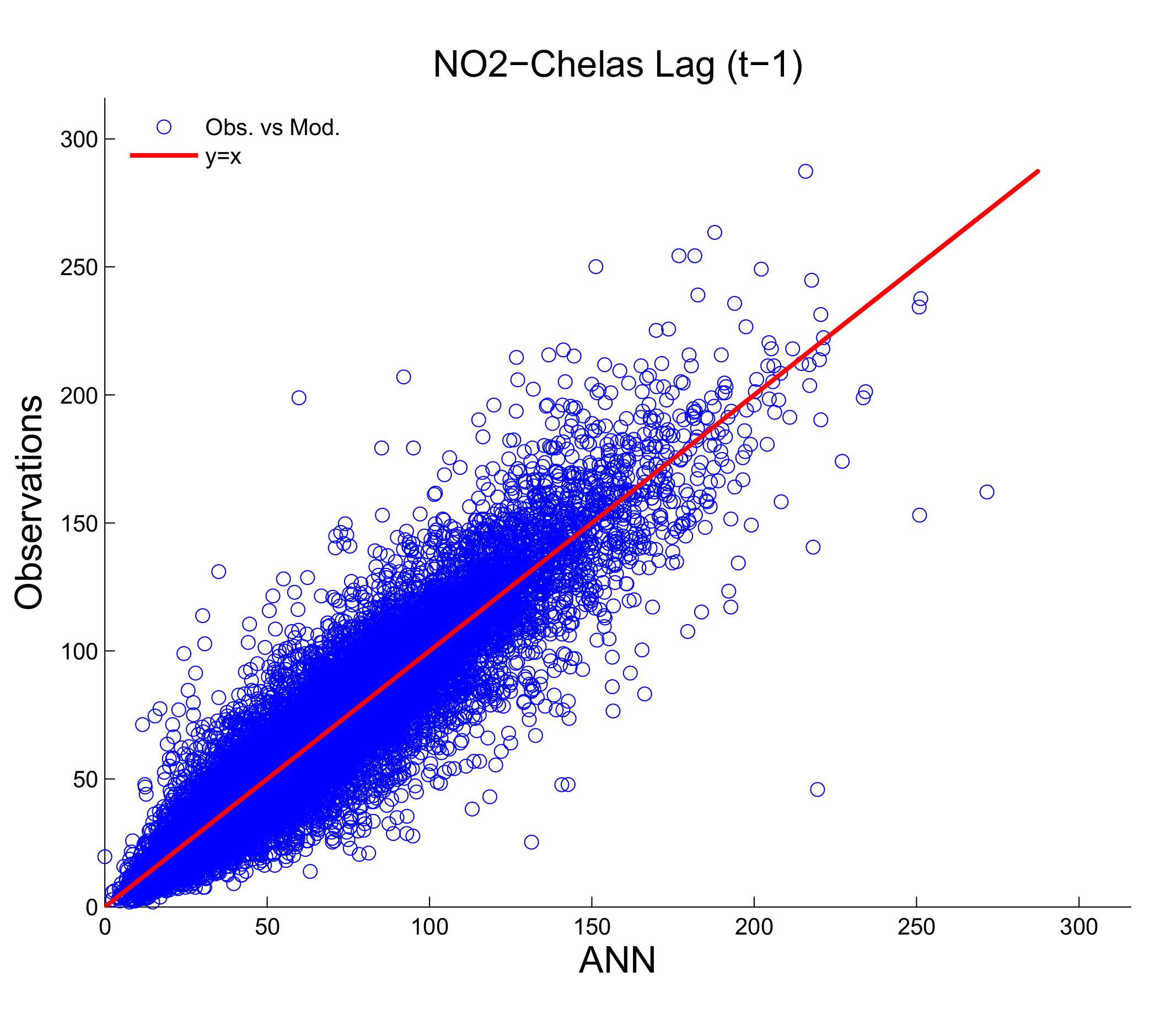}
 \caption{\protect 
           Scatter plot between observations and predicted values
           from the ANN for each station separately.
           The ANN is trained by
           the input variables selected by forward stepwise regression (FSR) elaborated in 
           Sec.~\ref{sec:inputdata}. In all cases the time lag is $1$ hour. }
\label{fig4}
\end{figure}
%%%%%%%%%%%%%%%%%%%%%%%%%%%%%%%%%%%%%%%%%%%%%%%%%%%%%%%%%%%%%%%%%%%%

%%%%%%%%%%%%Tabela de sumario dos trabalhos anteriores%%%%%%%%%%%%%%
\begin{table*}[t!]
%\footnotesize
\begin{center}
\begin{tabular}{|c|c|c|c|c|c|c|}
\hline
{\bf Lag} & {\bf Sta.} & {\bf Mod.} & {\bf PC} & {\bf Skillp} & 
{\bf RMSE}  & {\bf MI} \\
{\bf (h)} & &  & {\bf (\%)} & {\bf (\%)} & {\bf ($\mu g/m^3$)} &  \\\hline
\multirow{6}{0.2cm}{$\mathbf{1}$} & \multirow{3}{0.2cm}{{\bf C}} & $M_{s,FSR}$ & 92.8 (92.0) & 16.8 (54.5) & 12.9 (8.9) &  0.8 (0.8) \\ 
 &                                   & $M_{O}$ & 92.8 (91.8) & 16.1 (31.2) & 13.0 (8.7) & 0.8 (0.8)\\ 
 & & {\bf \% Dev} & {\bf 0 (-0.2)} & {\bf -4.1 (-42.8)} & {\bf +0.8 (-2.2)}  & {\bf 0 (0)} \\\cline{2-7}

 & \multirow{3}{0.2cm}{{\bf AL}} & $M_{s,FSR}$ & 92.8 (92.0) &17.2 (48.9) & 12.8 (16.8) &0.8 (0.6)\\ 
 &                                   & $M_{O}$ & 92.8 (92.0) &16.5 (46.1) &12.9 (16.7) &0.8 (0.5)\\ 
 & & {\bf \% Dev} & {\bf 0 (0)} & {\bf -4.1 (-5.7)} & {\bf +0.8 (-0.6)} &  {\bf 0 (-16.7)} \\\hline

\multirow{6}{0.2cm}{$\mathbf{3}$} & \multirow{3}{0.2cm}{{\bf C}} & $M_{s,FSR}$ & 76.8 (*) &33.2 (*) &22.2 (*) &0.5 (*)\\ 
 &                                   & $M_{O}$ & 76.8 (67.2) &33.2 (52.5) &22.2 (25.9) &0.5 (0.3)\\ 
 & & {\bf \% Dev} & {\bf 0 (*)} & {\bf 0 (*)} & {\bf 0 (*)} &  {\bf 0 (*)} \\\cline{2-7}

 & \multirow{3}{0.2cm}{{\bf AL}} & $M_{s,FSR}$ & 76.9 (71.8) &33.9 (56.3) &22.0 (30.6) &0.5 (0.3)\\ 
 &                                   & $M_{O}$ & 76.8 (71.2) &33.8 (53.8) &22.0 (30.8) &0.4 (0.3)\\ 
 & & {\bf \% Dev} & {\bf -0.1 (-0.8)} & {\bf -0.3 (-4.4)} & {\bf 0 (+0.7)}  & {\bf -20 (0)} \\\hline

\multirow{6}{0.2cm}{$\mathbf{6}$} & \multirow{3}{0.2cm}{{\bf C}} & $M_{s,FSR}$ & 65.2 (44.8) &47.1 (65.5) &26.3 (20.2) &0.3 (0.2) \\ 
 &                                   & $M_{O}$ & 65.0 (46.0) &47.0 (59.5) &26.3 (19.2) &0.3 (0.2)\\ 
 & & {\bf \% Dev} & {\bf -0.3 (+2.7)} & {\bf -0.2 (-9.1) } & {\bf 0 (-5.0)} &  {\bf 0 (0)} \\\cline{2-7}

 & \multirow{3}{0.2cm}{{\bf AL}} & $M_{s,FSR}$ & 65.4 (57.4) &47.8 (67.3) &26.0 (35.6) &0.3 (0.2)\\ 
 &                                   & $M_{O}$ & 65.2 (57.8) &47.5 (66.9) &26.1 (35.1) &0.3 (0.2)\\ 
 & & {\bf \% Dev} & {\bf -0.3 (+0.7)} & {\bf -0.6 (-0.6)} & {\bf +0.4 (-1.4)} &  {\bf 0 (0)} \\\hline

\multirow{6}{0.2cm}{$\mathbf{12}$} & \multirow{3}{0.2cm}{{\bf C}} & $M_{s,FSR}$ & 62.2 (39.0) &49.1 (60.4) &27.2 (20.9) &0.3 (0.1)\\ 
 &                                   & $M_{O}$ & 61.7 (38.4) &48.6 (60.0) &27.3 (20.9) &0.3 (0.1)\\ 
 & & {\bf \% Dev} & {\bf -0.8 (-1.5)} & {\bf -1.0 (-0.7)} & {\bf +0.4 (0)} &  {\bf 0 (0)} \\\cline{2-7}

 & \multirow{3}{0.2cm}{{\bf AL}} & $M_{s,FSR}$ & 63.3 (50.4) &50.4 (66.1) &26.6 (37.9)&0.3 (0.2)\\ 
 &                                   & $M_{O}$ & 62.7 (52.0) &49.8 (66.1) &26.8 (37.3) &0.3 (0.2)\\  
 & & {\bf \% Dev} & {\bf -0.9 (+3.2)} & {\bf -1.2 (0)} & {\bf +0.8 (-1.6) } &  {\bf 0 (0)} \\\hline

\multirow{6}{0.2cm}{$\mathbf{24}$} & \multirow{3}{0.2cm}{{\bf C}} & $M_{s,FSR}$ & 60.1 (45.8) &33.9 (54.9) &27.7 (19.7) &0.2 (0.2)\\ 
 &                                   & $M_{O}$ & 56.8 (56.4) &29.9 (40.7) &28.5 (18.1) &0.2 (0.2)\\ 
 & & {\bf \% Dev} & {\bf -5.4 (+23.1)} & {\bf -11.8 (-25.9)} & {\bf +2.9 (-8.1)}  & {\bf 0 (0)} \\\cline{2-7}

 & \multirow{3}{0.2cm}{{\bf AL}} & $M_{s,FSR}$ & 59.9 (61.1) &33.8 (55.1) &27.6 (34.2) &0.2 (0.2)\\ 
 &                                   & $M_{O}$ & 56.8 (62.7) &30.0 (55.3) &28.3 (33.3) &0.2 (0.2)\\ 
 & & {\bf \% Dev} & {\bf -5.2 (+2.6)} & {\bf -11.2 (+0.4)} & {\bf +2.5 (-2.6)} &  {\bf 0 (0)} \\\hline
\end{tabular}
\caption{\protect 
           Comparative overview of the efficiency and performance 
           for ANN according to two of the  models described in 
           Sec.~\ref{sec:inputdata}. To evaluate the efficiency
           and performance we use the Pearson coefficient (``PC''),Eq.~(\ref{PC}), 
           the skill against persistence (``Skillp''), 
           Eq.~(\ref{skillpers}), the root mean square error (''RMSE''),Eq.~(\ref{RMSE}), and the mutual
           information (``MI''), Eq.~(\ref{mi}). 
           The deviations (``Dev'')  give the relative improvement (or deterioration) of the optimized model  $M_O$ vs. $M_{S,FSR}$.
           As an overall conclusion, one can state that the inclusion of the  stochastic variable 
           produces similar results than the standard model, using much less input variables. 
           In parenthesis one shows the result obtained through independent 
           validation (see text).
(*) Not sufficient data to forecast.}
\label{fig_tabela_resultados}
\end{center}\end{table*}
%%%%%%%%%%%%%%%%%%%%%%%%%%%%%%%%%%%%%%%%%%%%%%%%%%%%%%%%%%%%%%%%%%%%

%%%%%%%%%%%%%%%%%%%%%%%%%%%
\section{Stochastic variables as optimal input for neural 
networks}
\label{sec:results}

In the previous section we concluded that one of our transformed 
variables is more correlated with the target than most of the
meteorological and air quality variables. 
Collecting only variables with an equal or higher correlation than 
the one observed for $V_O^+$, one obtains the model $M_{O}$, which
must now be compared with the standard model $M_{S,FSR}$ in its predictive
power.
In this section we present such comparison between both models.
Using different measures of predictive power, we conclude that, 
in general, the optimal model most frequently evidences the same 
predictive power as model $M_{S,FSR}$. Results are summarized in 
Tab.~\ref{fig_tabela_resultados}. 

The overall conclusion is that, despite the significant decrease of
input data, the predictive power of model $M_O$ is not worse than the
one of the models $M_{S,FSR}$.

To evaluate the efficiency and performance of our method four other
quantities are computed.
The first such quantity is the well-known Pearson correlation coefficient
(``PC'')\cite{Pearson},
\begin{equation}
\hbox{PC} = \frac{ \sum_{i=1}^N(y_i-\bar{y})(o_i-\bar{o}) }
           {\left[ \sum_{i=1}^N(y_{i}-\bar{y})^2\sum_{i=1}^N(o_{i}-\bar{o})^2 \right]^{1/2}} ,
\label{PC}
\end{equation}
where $y_i$ denotes the respective model forecast at time $i$ and $o_i$ denotes the real observed values at time $i$. 

From Tab.~\ref{fig_tabela_resultados} one sees that the difference
in the Pearson correlation between both models, $M_{S,FSR}$ and $M_{O}$,
is almost nonexistent, except for the largest time-lag, namely $t-24$. The PC results for the independent sample (2006) are relatively lower than for the calibration-validation period as expected, with the exception of the PC for AL (t=24 hours).

Another important quantity 
is the skill against persistence (``Skillp'') which can be 
interpreted as the percentage of improvement that our model can provide 
when compared with the persistence model, i.e.~the forecast for a given hour  
is the observed value of the previous time lag, i.e., 1h, 3h, 6h, 12h or 24h hours before. It is given by
\begin{equation}
\hbox{Skillp} = \frac{\frac{1}{N}\sum_{i=1}^N(y_i-o_i)^2-\frac{1}{N-1}\sum_{i=1}^N(\hat{y}_{i}-o_i)^2}{\frac{1}{N-1}\sum_{i=1}^N(\hat{y}_{i}-o_i)^2}\, ,
\label{skillpers}
\end{equation} 
where $\hat{y}_{i}$ is the value of the time series variable $y$  1, 3, 6, 12 or 24 hours before.

From Tab.~\ref{fig_tabela_resultados} one sees that the decrease %difference
in the skill against persistence between both models, $M_{S,FSR}$ and $M_{O}$,
is not significant ($\lesssim 4\%$) for short time-lags except for the largest time-lag, namely $t-24$. The skill against persistence results for the independent sample (2006) are higher than for the calibration-validation period for all the time lags.

Another quantity is the root mean square error (``RMSE'') 
which represents the difference between the pairs of forecast and 
observation values and is given by
\begin{equation}
\hbox{RMSE} = \sqrt{\sum_{i=1}^N(y_i-o_i)^2}.
\label{RMSE}
\end{equation}
While for shorter time-lags there is almost no difference between both
models,  a slight increase ($\lesssim 3\%$) of RMSE is observed
in the optimal model for the t=24 time lag. The RMSE for the independent sample are usually lower for the $M_{O}$, except for the AL t=3 case.

Finally, the mutual information evaluates the dependence between
observations and predictions and is given by

\begin{equation}
\hbox{MI} = \sum_{y\in Y_{bins}} \sum_{o\in O_{bins}} p(y,o) \log{\left ( \frac{p(y,o)}{p(y)p(o)} \right )} 
\label{mi}
\end{equation}
where $p(y,o)$ is the joint probability of observations and predictions
for the same time-steps and $p(y)$ and $p(o)$ are the corresponding
marginal distributions, computed within the range of admissible values,
$Y_{bins}$ and $O_{bins}$, properly discretized. With one single exception 
in the 2002-2005 period and one in the 2006 period, no deviations are observed when comparing the mutual information in both 
models.

Enabling a significant reduction of input data - less than one half -
and simultaneously presenting a predictive power at least as good as
the one given by the standard model, one can conclude that the model 
$M_{O}$ using our stochastic may be seen as a good alternative for
establishing ANN models in air quality prediction.
Figure \ref{fig4} shows for both Chelas and Avenida da Liberdade
the scatter plot between observations (\no\ measurements) and
the corresponding predictions from a ANN model training with
the optimized model $M_{O}$, indicating a consistently good agreement.

%%%%%%%%%%%%%%%%%%%%%%%%%%%
\section{Conclusions}
\label{sec:conclusions}

In this paper we applied a method for deriving eigenvariables in
systems of coupled stochastic variables, which were then used as input 
variables to considerably reduce the amount of input data needed for training
the ANN, typically for than a factor of two and for large time-lags by
a factor of ten. The predictive power is maintained. Indeed,
the introduction of the stochastic variables as input data for
training the ANN model allows to preserve the predictive power with considerable less input information.

This is because the variable incorporates temporal correlations between 
independent and spatially separated monitoring stations.

In the particular context of atmospheric environment, the reduction of the amount of input data, optimizes the ANN 
models in the sense that enables faster predictive outcome without
changing significantly the predictive power. Thus, our stochastic variables together with the findings of this
study can be taken as a first step for improving alarm systems, making 
them more efficient in predicting periods of lower levels in the air
quality.
Moreover, being a general numerical procedure for any given set of measurements,
our finding can be easily adapted to other ANN models in weather or
geophysical forecast. An extension of this work to take into account the 
correlations between a higher number of measurement stations is planned.

%%%%%%%%%%%%%%
\section*{Acknowledgments}

The authors thank DAAD and FCT for financial support through 
the bilateral cooperation DREBM/DAAD/03/2009.
FR (SFRH/BPD/65427/2009) and PGL ({\it Ci\^encia 2007})
thank Funda\c{c}\~ao para a Ci\^encia e a Tecnologia
for financial support, also with the support 
Ref.~PEst-OE/FIS/UI0618/2011.
The authors would like to acknowledge Ag\^encia Portuguesa do Ambiente and European Centre for Medium Weather Forecast for providing the environmental and meteorological data, respectively. 

%%%%%%%%%%%%%%%%%%%%%%%%%%%%%%%
%%% BIBLIOGRAPHY %%%%%%%%%%%%%%
%\end{linenumbers}

\bibliographystyle{elsarticle-num}

%%%%%%%%%%%%%%%%%%%%%%%%%END BIBLIOGRAPH%%%%%%%%%%%%%%%%%%%%%%%%%%%%%

\end{document}